# Fat and fibrosis as confounding cofactors in viscoelastic measurements of the liver

S S Poul[1], K J Parker[2]

[1]Department of Mechanical Engineering, University of Rochester, 235 Hopeman Building, Box 270132, Rochester, NY USA

[2]Department of Electrical and Computer Engineering, University of Rochester, 724 Computer Studies Building, Box 270231, Rochester, NY 14627, USA

Email:  kevin.parker@rochester.edu

**Abstract**

Elastography provides significant information on staging of fibrosis in patients with liver disease and may be of some value in assessing steatosis. However, there remain questions as to the role of steatosis and fibrosis as cofactors influencing the viscoelastic measurements of the liver tissues, particularly shear wave speed (SWS) and shear wave attenuation (SWA). In this study, by employing the theory of composite elastic media as well as two independent experimental measurements on fat-in-gelatin phantoms and also finite element simulations, it is consistently shown that fat and fibrosis jointly influence the SWS and SWA measurements. At a constant level of fat, fibrosis stages can influence the SWA by factors of 2-4. Moreover, the rate of increase in SWA with increasing fat is strongly influenced by the stages of fibrosis; softer background cases (low fibrosis stages) have higher rate of SWA increase with fat than those with stiffer moduli (higher fibrosis stages). Meanwhile, SWS results are influenced by the presence of fat, however the degree of variability is more subtle.  The results indicate the importance of jointly considering fat and fibrosis as contributors to SWS and SWA measurements in complex liver tissues and in the design and interpretation of clinical trials.

# 1. INTRODUCTION

Nonalcoholic fatty liver disease (NAFLD) spans a range of liver problems from simple steatosis, to early stages of fibrosis, to combination of steatosis and fibrosis, to fibrosis at advanced stages, and cirrhosis. Its prevalence is approximately 30% of the general populations in the United States and European countries which makes it one of the growing health concerns in the world. NAFLD develops initially (or is triggered) by an accumulation of lipid in the liver hepatocyte, greater than approximately 5%. Early diagnosis of NAFLD at the simple steatosis and early fibrosis stages could allow for treatment to reverse the disease process before it results in irreversible pathological damage to the liver (Ozturk *et al.*, 2018). The gold standard for diagnosing these conditions is the liver biopsy, which is invasive and uncomfortable for patients and also relies on data from a small sample of the liver tissue which might not be representative of the entire liver (Angulo, 2002; Chalasani *et al.*, 2018; Haga *et al.*, 2015).

Ultrasound (US) elastography techniques are non-invasive and affordable alternatives to biopsy and have drawn considerable attention for the prognosis and monitoring of histological changes to the liver during treatment (Parker *et al.*, 2010; Palmeri *et al.*, 2011; Barry *et al.*, 2012; Friedrich-Rust *et al.*, 2012; Nightingale *et al.*, 2015; Nenadic *et al.*, 2016; Langdon *et al.*, 2017; Barr, 2018; Parker *et al.*, 2018a; Ormachea *et al.*, 2019; Sharma *et al.*, 2019; Gesnik *et al.*, 2020). These studies aimed to characterize tissue properties and distinguish normal tissue from diseased tissue by correlating variation in measured biomechanical parameters with pathological changes.

In an ideal world, some ultrasound tissue characterization parameters could be derived which would produce a simple, monotonic increase with specific pathology and which would be largely independent of other cofactors or conditions. For example, ideally the shear wave speed (SWS) of liver tissue would increase monotonically with increasing fibrosis in a simple, sensitive, and accurate fashion, not influenced by other factors. Unfortunately, the role of cofactors can be major, and so various groups have attempted to mitigate or at least account for their roles (Ferraioli *et al.*, 2018). For example, a clinical study of patients with varying degrees of steatosis and fibrosis was reported by Petta *et al*. (2017). In this study, the correlation of the liver stiffness measurement (LSM), and the controlled attenuation parameter (CAP), (a proprietary ultrasound attenuation measurement), was investigated where steatosis and fibrosis coexist. It was shown that for livers where CAP is high, the degree of fibrosis is overestimated by LSM, and this results in an increase in false positives in the diagnosis of liver fibrosis.

As more and more measurements related to US and elastography parameters become available on commercial scanners, the role of cofactors must be carefully considered (Parker *et al.*, 2018a; Sharma *et al.*, 2019; Mikolasevic *et al.*, 2016). The roles of fibrosis and steato-fibrotic conditions on shear wave attenuation (SWA) measurements are less investigated and previous studies have mainly focused on the *acoustic* attenuation coefficient which is associated with the decay in the longitudinal compressional waves (Lin *et al.*, 1988). There was early disagreement in the results reported in the literature regarding the role of fibrosis on acoustic attenuation coefficients (Suzuki *et al.*, 1992; Afschrift *et al.*, 1987). Today, to the best of our knowledge, few studies have examined the role of fibrosis and steato-fibrosis on the *shear* wave attenuation where the results separate out the effect of the cofactors. Deffieux *et al*. (2015), in a study to investigate the effect of the viscosity on steatosis and fibrosis staging, reported no correlation between steatosis and viscosity.

Thus, two important clinical questions emerge in parallel: when we measure SWS in an attempt to gauge fibrosis, does the presence of fat (steatosis) confound or vary the results?

Similarly, when we measure SWA in an attempt to gauge the accumulation of fat, do varying degrees of fibrosis confound or alter the result?

In a naïve view, SWS would simply increase with fibrosis, while SWA would simply increase with the amount of fat accumulating in a steatotic liver. However, in reality the two conditions are confounding cofactors which need to be understood jointly. We address this issue by determining the cofactors' effects utilizing four independent views:

- From the theory of composite elastic media.
- From experimental stress relaxation measurements on fat-in-gelatin phantoms.
- From SWS and SWA measurements in fat-in-gelatin phantoms.
- From finite element (FE) simulations of shear waves in fatty livers.

These differing assessments lead to similar conclusions about the importance of fat and fibrosis as cofactors in liver elastography and are detailed in the following sections. The importance of these cofactors for stratifying clinical trials is a practical consequence of these findings.

## 2. THEORY

In the development of fibrosis, the shear modulus of liver typically increases. For a viscoelastic medium, the shear modulus is a complex parameter which is frequency-dependent and relates to the stiffness of the medium and the speed of wave propagation. When a shear wave propagates through a viscoelastic material, its two important propagation characteristics, SWS and SWA, depend on the complex shear modulus $G_c$ or the complex wave number $\hat{k}$ of the underlying material as following:

$$\hat{k} = \frac{\omega}{\sqrt{\frac{G_c(\omega)}{\rho}}} = \beta(\omega) - j\alpha(\omega) = \frac{\omega}{c_{ph}(\omega)} - j\alpha(\omega), \tag{1}$$

where $c_{ph}(\omega)$, $\beta(\omega)$, and $\alpha(\omega)$ are the phase velocity, the real part of the wavenumber, and the attenuation, respectively, all depending on the frequency $\omega$ and the density $\rho$ of the material (Carstensen *et al.*, 2008; Vappou *et al.*, 2009; Carstensen and Parker, 2014; Kazemirad *et al.*, 2016). Solving for $c_{ph}(\omega)$ and $\alpha(\omega)$ similar to the derivation of (Parker *et al.*, 2018b; Zvietcovich *et al.*, 2019), we have:

$$G_c(\omega) = G_{stor}(\omega) + jG_{loss}(\omega) \tag{2}$$

$$c_{ph}(\omega) = \sqrt{\frac{2|G_c(\omega)|}{\rho}} \left( \frac{|G_c(\omega)| + G_{stor}(\omega)}{|G_c(\omega)|} \right)^{-\frac{1}{2}} = |G_c(\omega)| \sqrt{\frac{2}{\rho(|G_c(\omega)| + G_{stor}(\omega))}} \tag{3}$$

$$\alpha(\omega) = \omega \sqrt{\frac{\rho}{2|G_c(\omega)|}} \left( \frac{|G_c(\omega)| - G_{stor}(\omega)}{|G_c(\omega)|} \right)^{\frac{1}{2}} = \frac{\omega}{|G_c(\omega)|} \sqrt{\frac{\rho(|G_c(\omega)| - G_{stor}(\omega))}{2}} \tag{4}$$

## 2.1 Composite theory

Steatotic liver tissue is characterized by microvesicular and macrovesicular accumulation of lipid vacuoles in the hepatocytes. Our approach is to model the simple steatotic liver as a composite medium with fat droplets considered as spherical inclusions distributed in a background material characteristic of the normal liver properties. In doing so, we can employ the theory proposed by Christensen (1969) and expanded by Lakes (1999) to model the fatty liver as a composite medium. Considering a normal liver with shear modulus $G_1(\omega)$, and fat inclusions with shear modulus $G_2(\omega)$ distributed in the normal liver with a small volume fraction of $V_2$, the simple steatotic liver will have a shear modulus of $G_c(\omega)$:

$$G_c(\omega) = G_1(\omega) \cdot \left\{ 1 - \frac{5\{G_1(\omega) - G_2(\omega)\}V_2}{3G_1(\omega) + 2G_2(\omega)} \right\} \quad (5)$$

with the assumption of a nearly incompressible medium consistent with normal tissues having a Poisson's ratio of $v_1 \sim 0.5$. This equation is valid for small volume fractions $V_2$ (and less than 0.5) and models a progressive departure from the properties of $G_1(\omega)$ as $V_2$ increases from zero. To model the normal liver $G_1$, we can employ the power law behavior using the Kelvin-Voigt fractional derivative (KVFD) model as follows:

$$G_1(\omega) = G_0(j\omega)^a = G_0 \cdot \omega^a \left\{ \cos\left(\frac{a\pi}{2}\right) + j\sin\left(\frac{a\pi}{2}\right) \right\}, \quad (6)$$

where $a$ is the power law parameter and $G_0$ is a constant. Moreover, we can model the fat inclusions as a viscous oil fluid with the viscosity of $\eta$ as a simple dashpot element with the shear modulus of:

$$G_2(\omega) = \eta \cdot j\omega \quad (7)$$

With the help of eqn (1) - (7) and our assumptions about fat being primarily a lossy term, we can now make some general statements about the interplay of factors. In practice, $G_1$ is in the range of 1 kPa for normal livers and dominates the $G_{stor}$ (real modulus) term, whereas $G_2$ is from fat inclusions that we model as a purely viscous material which contributes to the imaginary part of the modulus. Let us assume that increasing amounts of fibrosis create a progressively higher storage modulus $G_{stor}$ in eqn (2). In that case, $c_{ph}$ in eqn (3) will increase monotonically and directly as both $G_{stor}$ and $|G|$ increase. However, $\alpha$ will decrease because of the subtraction term in eqn (4). Now if fat is added in increasing amounts, which makes the volume fraction $V_2$ in equation (5) increase, the imaginary component $G_{loss}$ will increase according to eqn (7). In that case, in the "simple" range $V_2$ is small and $G_{stor}$ dominates initially; then as fat is added, $\alpha$ is increased through the increasing result of the subtraction term in eqn (4), and the material is actually softened by the addition of fat, resulting in a lower $c_{ph}$. As will be shown in the next sections, the accumulation of small amounts of fat in a fibrotic liver produces a slight decrease in SWS, this is easily disguised by other sources of variability. However increasing stiffness (fibrosis) creates a very strong drop in attenuation given a fixed amount of fat.

## 3. METHODS

To experimentally assess the role of fat and fibrosis as cofactors on the SWS and SWA measurements, two independent measures are employed to assess eight different viscoelastic phantoms. Separately, FE simulations are implemented to provide an independent test of the composite model. In this section, the details of experiments and the simulations are presented.

### 3.1 Phantom preparation

Eight different viscoelastic tissue-mimicking phantoms were made using a combination of gelatin powder, sodium chloride (NaCl), and agar in 900mL of degassed water forming the base mixture, and castor oil used for the inclusion. The portion of each ingredient is listed in **Table 1**. Four phantoms have 18% castor oil and four others have 2% castor oil, based on four different gelatin percentages of 3%, 4%, 5% and 6%.

Table 1 Portion of ingredients used for making viscoelastic phantoms.

| Ingredient | Amount |
|---|---|
| Gelatin | 3% |
| | 4% |
| | 5% |
| | 6% |
| Castor oil | 18% |
| | 2% |
| NaCl | 0.9% |
| Agar | 0.15% |
| Surfactant | 40cc/l oil |

In order to make the oil-water solution stay more stable, first the base mixture and the castor oil were separately heated up to a temperature of approximately 65°C and then oil was added to the gelatin mixture slowly while stirring constantly using a magnetic stirrer. A specific amount of surfactant was also added slowly to the oil-in-gelatin mixture to help keep the small oil droplets (already formed) suspended in the mixture without being aggregated in the whole process, making a uniform and stable oil-in-gelatin mixture. The solution was then cooled down to almost 30°C before it was poured into a cylindrical mold. The latter process was done slowly to avoid creating small bubbles in the mixture. The cylinder was sealed and placed on a low-speed rotator (model 33B, Lortone, Inc, Mukilteo, WA, USA) for almost 5 hours to rotate uniformly, letting the mixture solidify without oil drops aggregating. The phantoms were left at a temperature of 4°C overnight to solidify. The following day, the phantoms were allowed to reach room temperature before any ultrasound scanning or mechanical testing was done.

**Figures 1(a) and (b)** show a sample cut of a 4% pure gelatin phantom and a viscoelastic phantom with 4% gelatin and 18% castor oil, respectively, for comparison. In **Figure 1(c)**, a magnified view of the viscoelastic phantom in **1(b)** is presented where we observe a uniform homogeneous distribution of small drops of castor oil within the gelatin phantom. Most drops appear to have a diameter of less than 0.5 mm according to the magnified view.

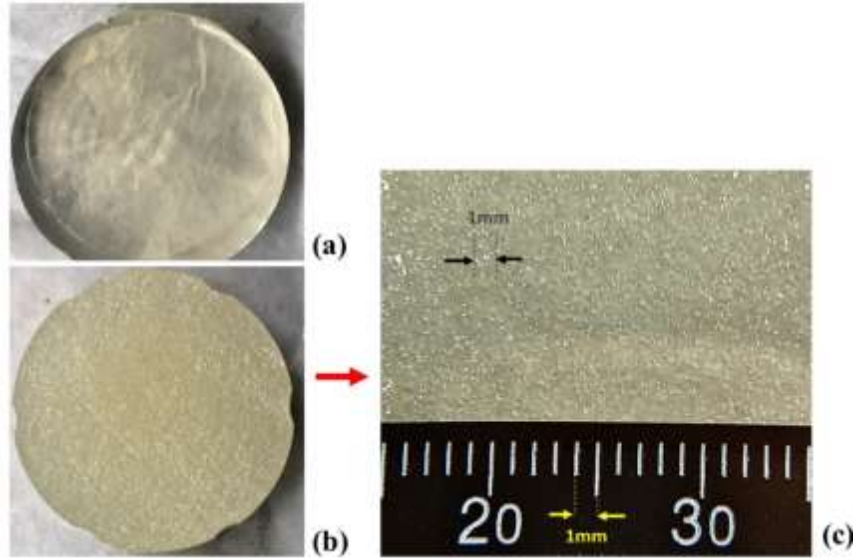

**Figure 1** Comparative structure of the phantoms: (a) pure elastic phantom with 4% gelatin, (b) 18% castor-oil-in-gelatin phantom with 4% gelatin, (c) magnified view of the phantom structure shown in (b) with the scale of millimeters for reference.

## 3.2 Ultrasound scanning

To obtain the mechanical properties of the viscoelastic phantoms and, therefore, the speed and attenuation of the shear waves propagating through the phantoms, a Samsung ultrasound scanner (model RS85, Samsung Medison, Seoul, South Korea) with a curved array transducer (model CAI-7A, Samsung Medison, Seoul, South Korea) was employed. It produced deformations that propagated as a shear wave in the phantom by applying radiation force excitation and then tracking the corresponding particle displacement. The center frequency of the transmit push beam is 2.5 MHz with a sampling frequency rate of 20 MHz. The SWS and SWA are calculated based on the theory presented in Parker *et al.* (2018c). The shear wave produced by the push pulse has a peak frequency in the range of 100 – 150 Hz in phantoms (Parker *et al.*, 2018b)

## 3.3 Stress relaxation test

Another independent test on oil-in-gelatin phantoms is the stress relaxation test that was employed to evaluate the properties of the viscoelastic phantoms. This compression test was done on 3-4 small cylindrical cuts with an average diameter of 20 mm and average height of 24 mm out of each cylindrical phantom, as shown in **Figure 2**. This test was done on the same day as the ultrasound scanning to ensure that the properties of the phantoms did not change due to dehydration or aging, and so that the comparison of the two modalities was more consistent. Using a Q-Test/5 machine (MTS, Eden Prairie, MN, USA) with a 5N load cell, a 5% strain was applied on each sample with a compression rate of 0.5 mm/s, and the relaxation test was done for approximately 500 s. Then, the stress variation in time for each sample was fitted to the stress vs. time relationship of the KVFD model, similar to the work by Zhang *et al.* (2008) to obtain the three coefficients $E_0$, $a$, and $\varsigma$ which also appear later in eqn (8). The complex Young modulus $E^*(\omega)$ as a function of frequency is obtained from eqn (8) using the three fitted coefficients as a last step. Assuming an incompressible material, the complex shear modulus is then calculated according to $G^*(\omega) = E^*(\omega)/3$.

$$E^*(\omega) = \left( E_0 + \varsigma \cos\left(\frac{\pi a}{2}\right) \omega^a \right) + j \left( \varsigma \sin\left(\frac{\pi a}{2}\right) \omega^a \right). \tag{8}$$

In this equation, $a$ is the power law parameter, $\varsigma$ is related to the viscous behavior of the material, and $E_0$ is an elastic modulus constant which is negligible for soft tissues and viscoelastic phantoms.

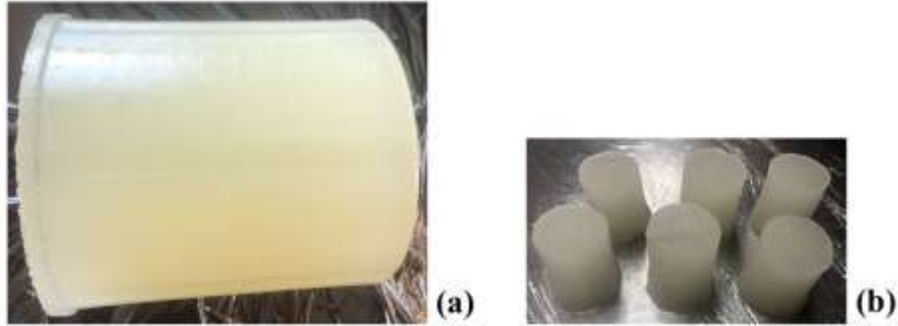

**Figure 2 (a)** A sample of a large cylindrical phantom (4% gelatin and 18% castor oil) **(b)** Small cylindrical cuts for the stress relaxation test.

## 3.4 Finite element simulation

Shear wave propagations through homogeneous and inhomogeneous media were numerically simulated using Abaqus/CAE 2019 (Dassault Systems, Vélizy-Villacoublay Cedex, France). The simulation domain is a 3D block with the $z$-direction as the propagation direction and the $x$ and $y$ as the lateral directions. The block is subjected to a three-cycle 150 Hz toneburst transient shear displacement excitation along the $y$-direction, where the displacement values are symmetric with respect to both *x* and *y* axes. A 3D schematic of the block, its orientation, and the excitation plane are depicted in **Figures 3(a)** and **(b)**.

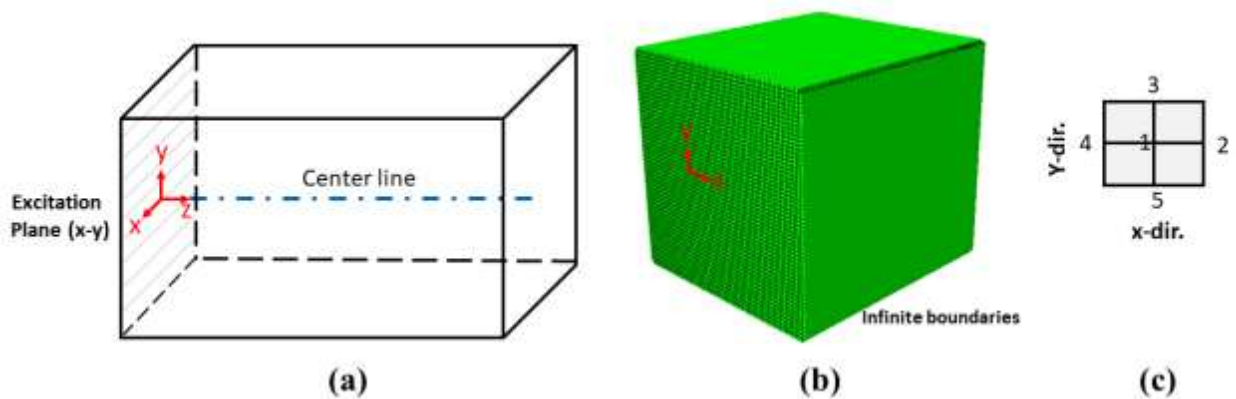

**Figure 3** (a) Schematic of 3D block orientation, excitation plane, and the propagation centerline. (b) Computational meshes of the domain with the infinite boundaries in the FE simulation. (c) Definition of the four neighboring points around each center point on the *z* axis, a few wavelengths apart.

The domain is meshed using 214,816 hybrid, quadrilateral linear elements (C3D8RH). The mesh size is refined to resolve the smallest wavelength in each simulation. In order to avoid the reflection of the incident wave from the boundaries back into the domain and to avoid the unwanted interference, infinite boundaries are defined around the domain to minimize the reflection. The simulation models approximately 50 ms of wave propagation in the computational domain based on the dynamic-implicit method.

For the inhomogeneous simulation, the inclusion material is distributed randomly, as single mesh elements within the background material of the 3D domain. The background material is modeled as an elastic material with a density of 998 kg/m$^3$ and a Poisson's ratio of 0.495. The inclusion is characterized as an almost purely viscous material using a Zener model with small $E_\infty$ and relatively high $E_1$. The displacement at a number of points along the centerline of the propagation direction ($z$) is calculated as well as four neighboring points around each $z$-location. The displacement at each $z$-location is the average of the displacement at that specific point and the displacement at the other four neighboring points. The arrangement of these neighboring points is illustrated in **Figure 3(c)**.

In order to simulate the simple steatosis condition, four different inclusion percentages of 6%, 12%, 18%, and 24% were implemented in Abaqus. Moreover, to simulate the effect of fibrosis and the base material stiffness level on the SWS and SWA parameters, five different background materials were set up in Abaqus. The stiffness levels used for modeling fibrosis stages in the simulations are based on the METAVIR scoring system which is selected based on peak of the probable values of SWS (stiffnesses) for the fibrotic livers presented in the statement by the Society of Radiologists in Ultrasound (Barr *et al.*, 2015). These five groups and their selected material SWS, which also represents the stiffness level, are reported in **Table. 2**.

**Table 2.** Background material SWS for simulating different fibrosis stages.

| Fibrosis score (METAVIR) | SWS (m/s) |
|:---:|:---:|
| F0 | 0.9 |
| F1 | 1.1 |
| F2 | 1.4 |
| F3 | 1.75 |
| F4 | 2.2 |

Considering the effect of fat and fibrosis simultaneously, 20 simulations in total were performed based on different fat inclusions and different background fibrosis (stiffness) stages. For each inhomogeneous simulation with the inclusion, to compensate for the effect of geometric spreading on the amplitude decay, a corresponding homogeneous elastic simulation is also performed at the same group velocity. Therefore, we can quantify the SWA as an exponential decay in peak amplitude, corrected for geometric spreading, as a function of how stiff the background is and how much fat inclusions accumulate within the medium.

## 4. RESULTS

### 4.1 Phantom experiments

The general trend in the stress relaxation curves at constant strain rate, as shown in **Figure 4**, is an increase in the stress level with increasing gelatin percentage when the castor oil inclusion amount is fixed. This trend is also observed in the value of the $\varsigma$ parameter in the KVFD model: $\varsigma$ increases significantly with increasing gelatin percentage. The KVFD power law parameter $a$ oscillates in a small range around 0.045 for all cases, and $E_0$ is also negligible as expected, for viscoelastic material behavior. The details are reported in **Table 3**.

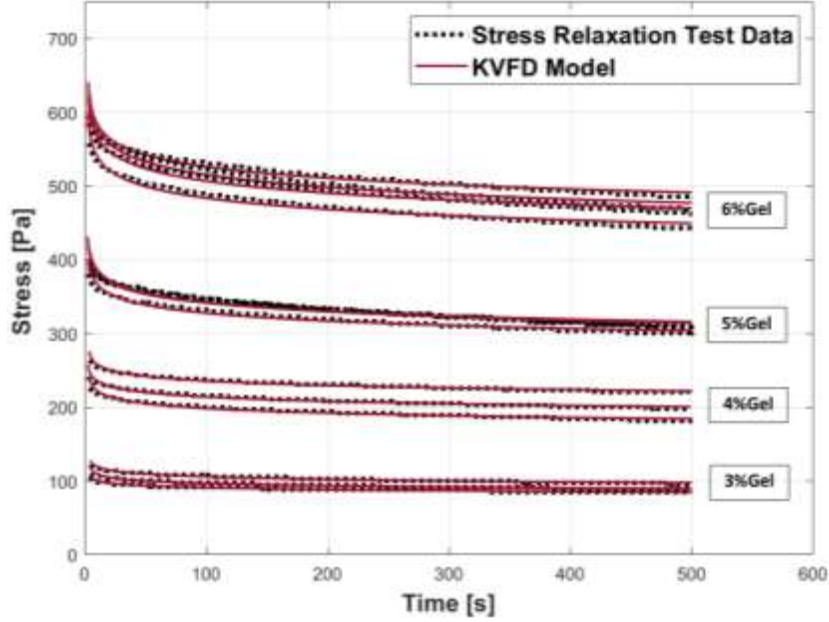

**Figure 4** Stress relaxation test for phantoms with 18% castor oil and different gelatin percentages: 3%, 4%, 5%, and 6%.

**Table 3.** KVFD parameters for each viscoelastic phantom.

|  | $E_0$ | $a$ | $\varsigma$ |
|---|---|---|---|
| 3% gelatin | 2.19E-05 | 0.046 | 2487 |
| 4% gelatin | 2.11E-04 | 0.045 | 5038 |
| 5% gelatin | 1.30E-04 | 0.049 | 8765 |
| 6% gelatin | 4.68E-05 | 0.045 | 12874 |

First, we utilized the two independent sets of results for SWS and SWA from both mechanical stress relaxation tests and ultrasound scans on viscoelastic phantoms and compared the two with the composite theory predictions. In employing the composite model for the theoretical estimation of the shear modulus of the phantom with an 18% castor oil inclusion $(G_c)$, the shear modulus of the background material $(G_1)$ is needed according to eqn (6). This is approximated by the shear modulus of a 2% castor oil phantom at the same background stiffness level. The reason for using the 2% results instead of pure gelatin (0% oil) is due to the observation that the addition of minimal castor oil drops with the surfactant and rotational processing may change the conformation of the gelatin background material itself. Therefore, the 2% castor oil is a sufficiently small amount of oil to represent the asymptotic approach of the composite properties to near zero inclusions.

In **Figures 5 (a) and (b),** the SWS and SWA are shown for different background stiffnesses (gelatin percentages) at 18% oil inclusion. The SWS and SWA are both calculated from two independent tests of: (i) ultrasound scan results used with the theory in Parker *et al*. (2018c), and (ii) the mechanical test results fit to the KVFD model at 150 Hz frequency. We find that results from both tests are consistent with the composite theory predictions for SWS as well as SWA. The SWS increases with the increase in background stiffness and for the SWA, the general trend is decreasing SWA with increasing background stiffness, an observation supported by theory and phantom experimental results. The ultrasound scan results, KVFD estimates, and theory predictions are shown as box plots, blue bars and dashed line, respectively.

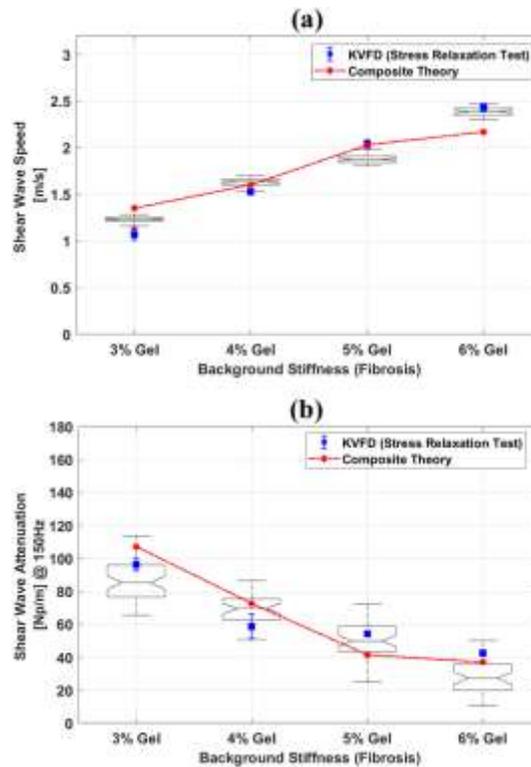

**Figure 5** Comparison of (a) SWS (b) SWA, at 18 % fat and different fibrosis stages for the composite theory vs. the stress relaxation test, and Samsung scan results shown as box plots.

A sample of B-scan and elastography images for the 4% and 6% gelatin phantoms both with the 18% castor oil inclusion are shown in **Figures 6(a) and (b)** with the average shear wave speed and attenuation coefficients.

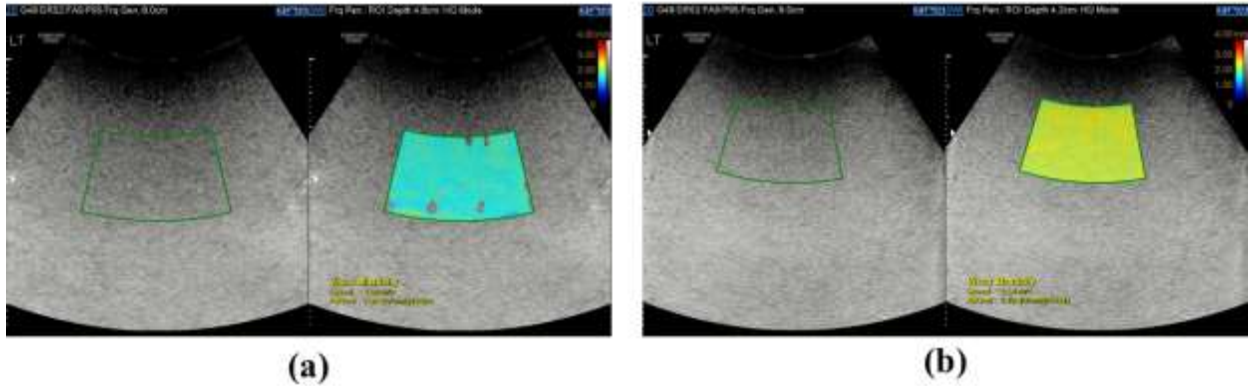

**Figure 6** Sample of a B-scan (left side) and elastography images (right side) of a **(a)** 4% gelatin phantom and **(b)** 6% gelatin phantom, both having 18% of castor oil inclusions.

## 4.2 Finite element simulation

Shear wave propagation results were evaluated from the FE simulations, and the SWS was obtained using the time-of-flight method. SWA was estimated from an exponential decay curve-fit after comparison against the geometric spreading in a corresponding elastic (non-attenuating) homogeneous medium of the same group velocity. The presence of inhomogeneous inclusions changes the wave front and also the displacement at nodal points.

**Figures 7(a) and (b)** show the propagating waveform of the homogeneous pure elastic and inhomogeneous 12% inclusions, respectively, both at F4 fibrosis (stiffness) stage. The presence of small inclusions in the background alters the wave front border, deviating from the uniform unperturbed front in the homogeneous simulation to a disturbed rippled front in the inhomogeneous case. Specifically, looking at the displacement field at the same locations along the propagation distance in **Figures 7(c) and (d)**, we see that the presence of the fat inclusions results in a decrease in the level of displacement in comparison to the homogeneous case.

Moreover, the homogeneous pure elastic case itself presents amplitude decay along the propagation direction, which is associated with the geometric spreading of the wave.

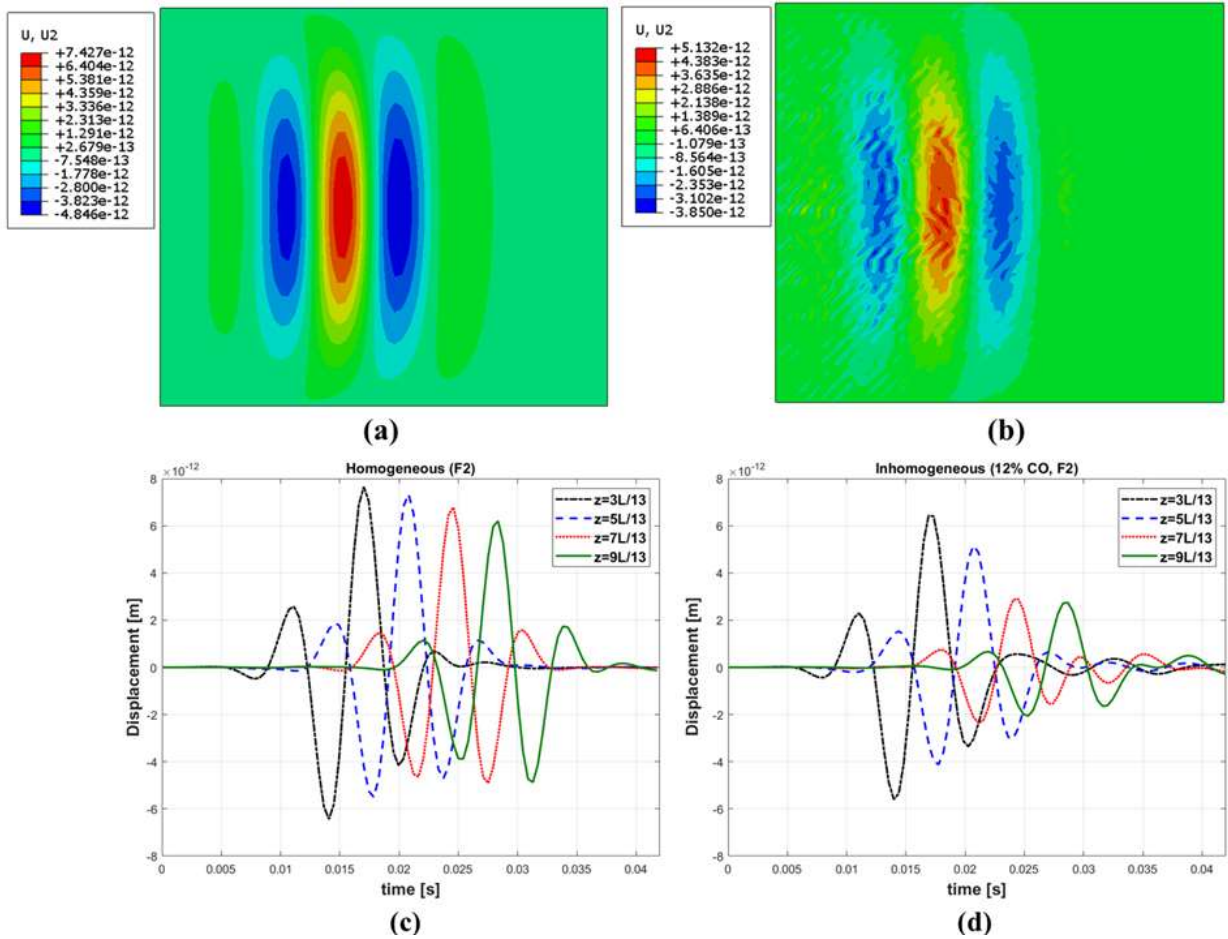

**Figure 7** Top row: the propagating waveform in the F simulation for (a) the homogeneous medium (b): the inhomogeneous medium. Bottom Row: Time evolution of displacement at four different locations along the *z*-direction in the FE simulations in the (c) homogeneous and (d) inhomogeneous medium. All cases are at the fibrosis (stiffness) level of F4. The inhomogeneous medium has 12% inclusions.

**Figures 8(a) and (b)** illustrate the comparison of composite theory with the simulations for SWS, respectively. These figures depict the elevated SWS with the advance in fibrosis (background stiffness) stage and also the reduced SWS with the development of higher fat content. SWA comparison of composite theory with FE simulations are also presented in **Figures 8(c) and (d),** respectively. The two plots indicate the decreased level of SWA at higher fibrosis stages and also the increased SWA as a function of higher steatosis score. The plots of **Figure 8** also indicate agreement between the theory and simulation for SWS and also the similar trend for SWA values.

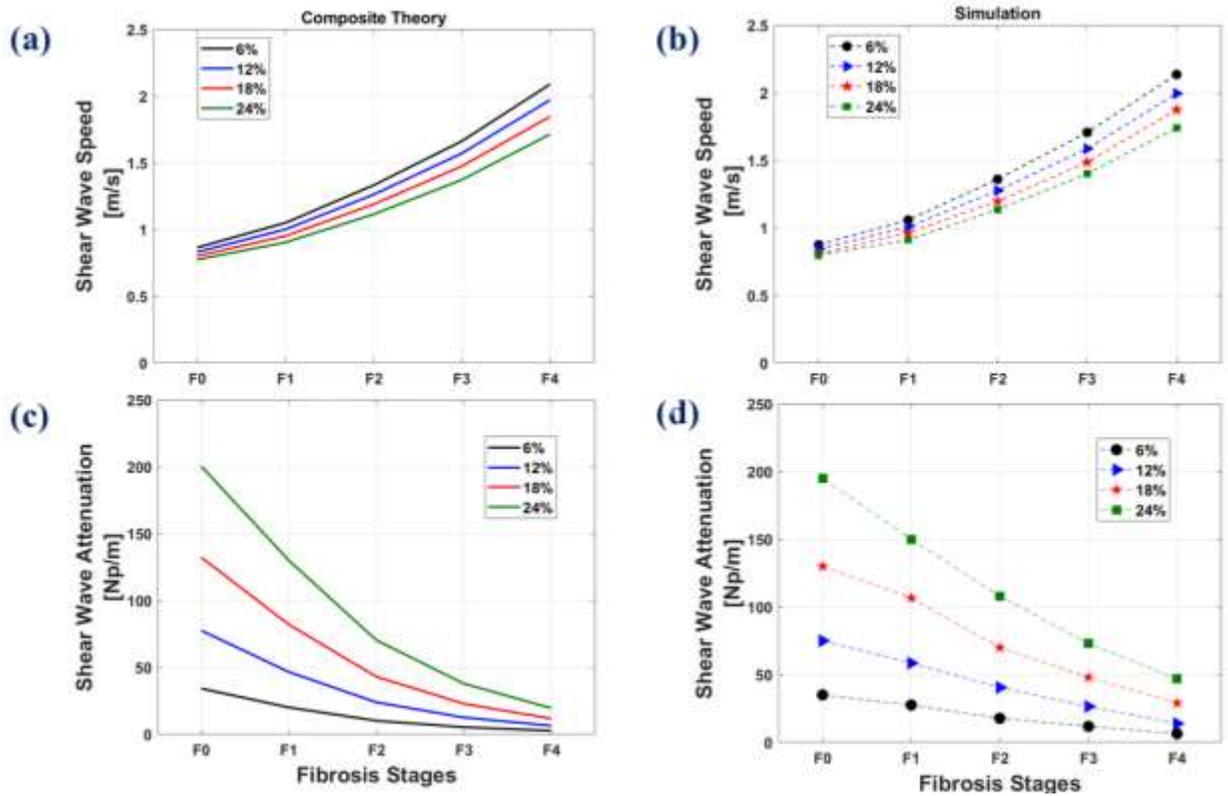

**Figure 8.** Composite theory vs. FE simulations at different fat percentages and different fibrosis stages Top row: SWS of **(a):** composite theory, **(b):** FE simulation. Bottom row: SWA of **(c):** composite theory, **(d):** FE simulations.

## 5. DISCUSSION

### 5.1 General trends

Good agreement was observed between the trends of results for SWS and SWA from three different estimates derived from ultrasound shear wave scanning, stress relaxation tests, and the composite theory, all supporting the importance of considering the cofactors of fat and fibrosis. These trends are also observed in the results from FE simulations for the two parameters of SWS and SWA; this further underscores the significance of the two factors. Fat accumulation in low volume percentage is a weak cofactor influencing (decreasing) SWS, however this effect will be frequency-dependent and so could be confusing when comparing different studies' results using different shear wave frequencies. However, at higher fat volume percentages, fat starts to decrease the SWS more significantly. On the other hand, baseline stiffness changes create a pronounced influence on SWA. This suggests the significance of considering these potential cofactors when interpreting the SWS and SWA measurements and correlating them with the histological conditions of tissues in diseases which have not yet been studied to the best of our knowledge.

In comparing the SWA of the composite theory and the FE simulations, although the trend is the same across different stiffness levels and different inclusion percentages, the SWA values from simulations are higher than that of theory. One of the important reasons behind that is the fact that the analytical solution in the composite theory is based on purely elastic theories in which the scattering phenomena are absent. But in numerical simulations and also experiments, some degree of scattering of shear waves is present. The wave scattering occurs when the wave propagates in an inhomogeneous medium with an impedance mismatch between the medium and the small inhomogeneities (Wu and Aki, 1985). This introduces an additional component of loss to the forward propagating wave and therefore the estimated SWA coefficient would be higher in simulations that incorporate scattering phenomena.

**5.2 Physics vs. statistics in clinical trials**

In elastography clinical trials, a population may be studied under broad inclusion criteria incorporating different degrees of liver fibrosis and steatosis. Frequently, a linear correlation fit of the metrics against an independent diagnostic assay is attempted. To look at the cofactors' roles (fibrosis and steatosis) on the SWA and SWS measurement, let us assume one patient is sampled for each of the 20 parameter pairs shown in the solid points of **Figures 8(b) and (d)** (five values of fibrosis, F0-F4; and four values of fat concentration for each fibrosis score). Since in clinical practice there are unavoidable errors in measurements of fat content and also shear wave propagation parameters, for more realistic measurements we added a proportional 10% Gaussian noise to both the SWA and fat inclusion percentage measurements in **Figure 8(d)** and also to both the SWS and background stiffness in **Figure 8 (b).** For this distribution of parameters, some simple linear correlation plots of SWA vs. percent fat are presented in **Figures 9(a) – (d)** and the correlation plot of SWS vs. fibrosis stage (stiffness) is shown in **Figure 9(e)** with the corresponding linear correlation fitting parameters reported in **Table 4** for our simulation results. For the SWA correlation with fat, the overall population of 20 cases are shown in **Figure 9(a)**, and while a trend to increasing SWA with increasing fat is observed, the correlation is poor, and the variability of data is pronounced.

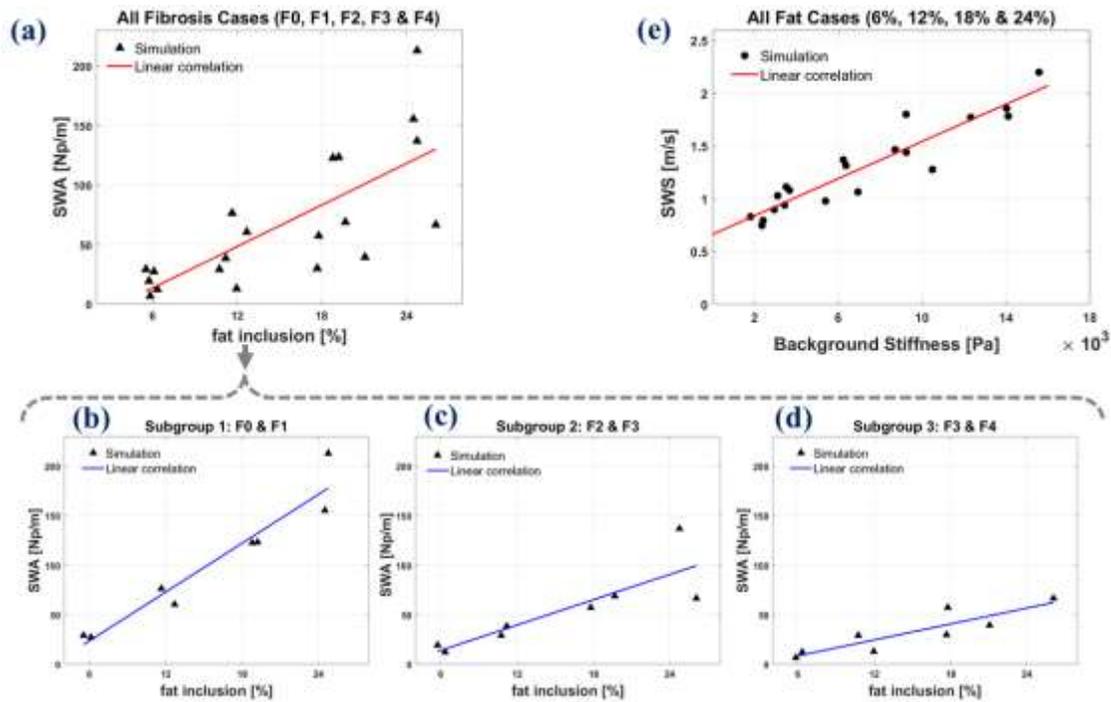

**Figure 9** Correlation of SWA with fat inclusion percentage incorporating **(a):** all fibrosis cases, **(b):** F0 and F1 cases as subgroup 1, **(c):** F2 and F3 cases as subgroup (2) and **(d):** F3 and F4 cases as subgroup 3. **(e):** Correlation of SWS with fibrosis (background stiffness) at all fat inclusion cases.

**Table 4**: Linear correlation details of the SWA with fat inclusion and the SWS with fibrosis level

| Cases | $R^2$ (goodness of fit) | Correlation slope $m$ (as in $y = mx + B$) |
|---|---|---|
| All fibrosis cases (F0, F1, F2, F3 & F4) | 0.555 | 5.82 (Np/m) / (% fat) |
| subgroup 1 (F0 & F1) | 0.924 | 8.21 (Np/m) / (% fat) |
| subgroup 2 (F2 & F3) | 0.723 | 4.26 (Np/m) / (% fat) |
| subgroup 3 (F3 & F4) | 0.785 | 2.68 (Np/m) / (% fat) |
| All fat cases (6%, 12%, 18% & 24%) | 0.874 | $8.81 \times 10^{-2}$ ($\frac{m/s}{kPa}$) |

However, when different subgroups of fibrosis stages are analyzed separately, the correlation plots are improved with enhanced $R^2$. Moreover, this correlation is affected by the level of fibrosis stages: the lower fibrosis stages have higher correlation slopes of SWA with increasing fat as shown in **Table 4.** For instance, subgroup 1 as the combination of F0 and F1 groups has the highest correlation slope and $R^2$. This stratification by the degree of fibrosis improves the tighter interpretation of SWA measurements. Looking at the correlation plot of SWS vs. fibrosis stage for the overall population in **Figure 9(e)**, we observe a trend of increasing SWS with increasing fibrosis. This plot has relatively less variation and spread of the SWS data due to the presence of different fat inclusion percentages. The value of $R^2$ as a measure of correlation goodness of fit in **Table 4** also supports this observation.

Finally, there are additional factors that could confound the interpretation of SWS and SWA measurements in complex liver tissues, for example inflammation, lesions, and vascular pathologies. These represent other cofactors that need to be modeled as influences on viscoelastic properties for a better overall judgement of measurements. Another important factor is the shear wave frequency. We have focused on shear waves near or at 150 Hz based on values recorded from push pulses (Parker *et al.*, 2018b; Ormachea and Parker, 2020, in press), however elastography using ultrasound, magnetic resonance (MR), and optical coherence tomography (OCT) can incorporate lower frequencies such as 50 Hz for large organs, or much higher frequencies of 1-2 kHz for small organs or structures. The linear dependence of viscosity on frequency in eqn (7) is a strong driver of the effect of fat, and this remains as a key parameter that requires further verification against the composite theory.

## 6. CONCLUSION

In this study, we find consistent results from composite theory, from two independent experimental measures, as well as FE simulations, all describing the role of steatosis and fibrosis as cofactors on SWS and SWA measurements. The results indicate that SWA and SWS are influenced by both the amount of fat and also the level of background stiffness. When the fat inclusion percentage is kept constant, the measured SWA will vary with the fibrosis stages by factors of 2-4. Furthermore, fibrosis stages have strong effects on the rate of change in SWA with respect to fat, i.e., cases with softer background show higher rate of change in comparison to the cases having stiffer background. On the other hand, the influence of fat on SWS is less dramatic and could easily be obscured in studies with significant measurement errors. The effect of accumulating fat is also a strong function of shear wave frequencies, so our examples must be understood to be representative of shear waves in the band around 150 Hz as produced by some systems' push pulses. The joint influence of fat and fibrosis can be considered within viscoelastic models, or can be simply minimized in practice by designing clinical trials so as to stratify research subjects' measurements into subgroups.

**Acknowledgments**

This work was supported by National Institutes of Health grant R21EB025290. We thank Samsung Medison Company Ltd. for the use of the scanner.